\documentclass{cimento}
\usepackage{graphicx}
\title{Raman Scattering of Metals to Very High Pressures}
\author{Alexander F. Goncharov\from{ins:x}\ETC,
Eugene Gregoryanz\from{ins:x}, Viktor V. Struzhkin\from{ins:x},
Russell J. Hemley\from{ins:x}, Ho-kwang Mao\from{ins:x}, Nabil
Boctor, \from{ins:x},
Eugene Huang\from{ins:z} } \instlist{\inst{ins:x} Geophysical
Laboratory and Center for High Pressure Research,Carnegie
Institution of Washington, 5251 Broad Branch Rd, N. W., Washington
D.C. 20015, USA
\inst{ins:z} Institute of Earth Sciences, 128 Academia Rd., Sec.
2, Nankang, Taipei, Taiwan 115} \PACSes{\PACSit{63.20Kr.,
62.50.+p,78.30.-j}
\PACSit{---.---}{\ldots}}
\begin{document}

\maketitle

\begin{abstract}


\end{abstract}

\section{Introduction}

The study of the effect of pressure on materials is fundamental to
a range of problems in physics (e.g., Ref. \cite{Hemley}). With
increasing pressure, the distances between atoms normally
decrease, which leads to the increase of interparticle
interactions.  Thus, by changing the pressure, one can tune the
physical properties of materials and observe corresponding changes
in the vibrational, electronic and magnetic excitations. The
gradual changes of interatomic distances and interaction in some
cases yield to radical transformations that are unique to high
pressure. The behavior of compressed materials can be studied
under static conditions to pressures above 300 GPa (or 3
megabars). The associated large decrease in volume provides a
wealth of opportunities to observe basic changes in bonding
character, magnetic or electronic structure or chemical state.
Such experiments have become possible with the development of
ultrahigh-pressure diamond-anvil cell techniques (e.g., Ref.
\cite{hemley97}). In particular, high-pressure spectroscopy
provides crucial and often unique information on bonding
properties and excitations of metals, semiconductors and
superconductors \cite{porter,gillet}. Ultrahigh-pressure research,
by its very nature, imposes substantial requirements on
experimental technique. The extremely small amount of the material
(e.g., down to picoliter volumes) brought to megabar pressures
demands extremely sensitive and versatile equipment. Thus, the
quality of information obtained critically depends on the
development of new techniques and improvement of the existing
methods.

Raman studies of metals is a challenging task because only a very
thin surface layer of the sample (skin-depth of $\leq$1000$\AA$)
is typically probed by the exciting laser radiation. This makes
the requirements for the experimental technique even more severe.
Raman studies in metals were began in 1966, with the pioneering
work of Feldman {\it et al}. \cite{feldman}. They were able to
detect very weak signals from phonon modes in Be and AuAl$_2$. The
first work on metals under pressure were published in 1992 by
Olijnyk \cite{olijnyk}. The high-pressure phases of Si and Ge were
studied to 50 GPa by Raman measurements of the phonon modes, and
the results were in reasonable agreement with theoretical
calculations \cite{needs,chang85,chang86}. Such studies have now
become more feasible, reliable, and definitive with the
development of new CCD array detectors and optical systems
\cite{williams,delhaye}, because they give tremendous advantage in
comparison to single-channel and diode-array techniques.

Raman spectroscopy of metals has been of growing interest, in part
because it is a very informative way of characterizing
high-temperature superconductors (including determination of the
superconducting gap; e.g., Refs. \cite{blumberg,hackl}). The
possibility of using Raman spectroscopy under pressure for studies
of these materials has been demonstrated \cite{goncharov93}.
Several applications of the technique to metals have been reported
\cite{olijnyk97,jephcoat,olijnyk2000}, but its more extensive use
has been limited in pressure range and by the need to detect very
weak signals. We applied new holographic transmission optics Raman
techniques \cite{tedesco} for high-pressure studies to obtain
substantially improved performance, which allowed us to extend
Raman measurements in metals to the megabar pressure range
\cite{merkel}.

\section{Experimental}

\subsection{Raman technique}

\begin{figure}
\includegraphics[width=6in]{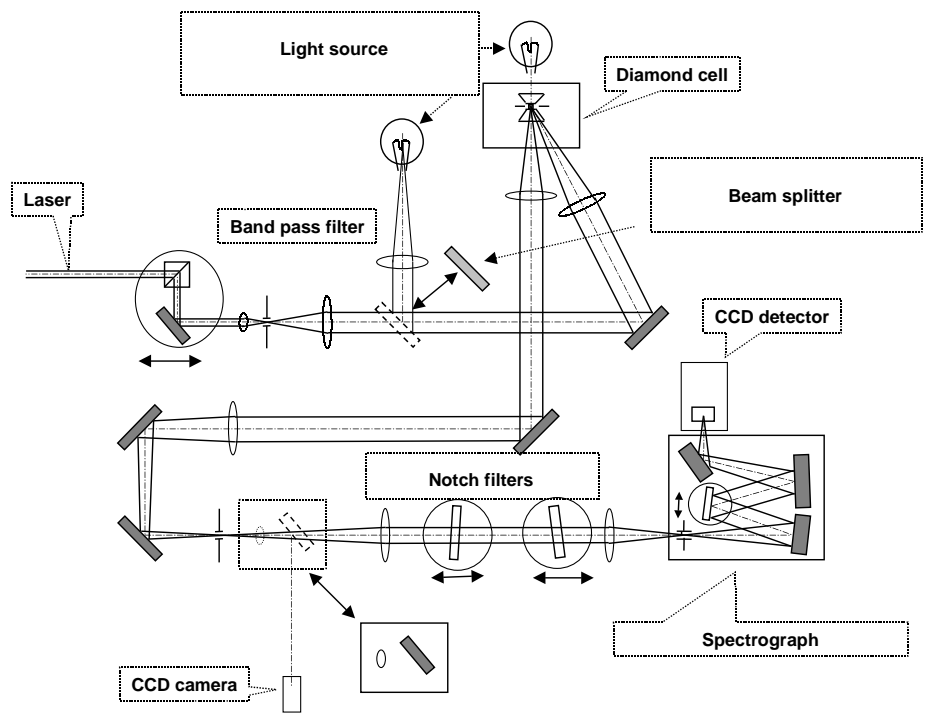}
\caption{Layout of an optical system used for the study of metals
under pressure.} \label{fig1}
\end{figure}

The recent improvements in Raman spectroscopy techniques stemmed
from the development of extremely sensitive array detectors (CCD)
and fast imaging spectrometers \cite{gillet,delhaye} coupled with
and incorporating holographic transmission optics. The latter has
already found a wide application as notch filters, which
efficiently block undesirable Rayleigh-scattered light in Raman
systems used for routine measurements \cite{yang}. This obviates
the need for cumbersome low-transmission double subtractive
monochromator filter systems for measurements above 50 cm$^{-1}$
and increases substantially the throughput of the system. The
obvious drawback of such systems in its commercial realization is
the need of a separate set of optics for each excitation
wavelength (e.g., lines of an Ar-ion laser).

In the case of high-pressure spectroscopy, flexible changes in
both the excitation wavelength and the spectral range are
essential because of the need to choose the optimum excitation for
both ruby fluorescence and Raman measurements. This requires
convenient tuning of the central wavelength of the laser bandpass
filter and the position of the notch of the laser rejection
filter. We have developed a simple universal design for
Raman/fluorescence measurements with capabilities of collecting
low- and high-resolution spectra and exciting spectra in different
geometries using a variety of laser wavelengths, which can be
easily selected by tuning holographic optics (fig. \ref{fig1}).
The spectral and spatial purity of the laser beam is greatly
improved by use of a bandpass filter (Kaiser Optical Systems) and
a spatial filter, which is also used as a beam expander. The
collimated laser beam is focused to the sample in either
backscattering geometry with a neutral or dispersive beamsplitter
\cite{patent} and the same long working-distance objective lens,
which is used for collecting the scattering radiation, or at an
oblique angle using an auxiliary long focal-length lens. The
scattered radiation collimated by the objective lens passes one or
two notch filters (Kaiser Optical Systems) and is focused into a
confocal pinhole, which is conjugated to the entrance slit of the
spectrograph, to reduce the unwanted fluorescence from the diamond
anvils. The best performances have been realized with the 460-mm
focal length f/5.3 imaging spectrograph (Jobin Yvon HR460)
equipped with two 1800 and 300 grooves/mm gratings on the same
turret. This feature allows one first to take a quick spectrum
over a wide spectral range (about 5000 cm$^{-1}$) and then study
regions of interest with higher accuracy and resolution. The
system can be tuned to other excitation frequencies by changing
the laser wavelength and rotating the holographic filters. One set
of holographic filters covers the spectral range about 70 nm
(notch filters) and 150 nm (bandpass filter), which is suitable
for most applications. Use of notch filters for laser radiation
rejection limits the spectral range available for Raman
measurements to $\Delta\nu>$50-100cm$^{-1}$ because the bandpass
of the notch filter cannot be made narrower with present
technology. If the low-frequency range is needed, then the
spurious laser radiation can be supressed by a double tunable
filter made with dispersive holographic transmission gratings
\cite{patent}.

Use of the double spatial filtering (one for the laser and one for
the signal) effectively suppresses laser plasma lines and unwanted
Raman/fluorescence signals (e.g., from diamond anvils). In this
configuration the laser spot can be made very tight (e.g., 2
$\mu$m and the depth of focus is substantially reduced. Thus, the
Raman signal from the diamond anvil is suppressed to a minimum;
the effects of pressure unhomogeneity can also be substantially
reduced \cite{goncharov99}. Diamond fluorescence can be further
reduced by using synthetic ultrapure diamonds anvils. This is
crucial for studies of very weak scatters (e.g., high-temperature
superconductors) and/or materials at very high pressures (above
200 GPa), where strong stress-induced fluorescence of diamond
anvils is a major obstacle for obtaining Raman/fluorescence
spectra of samples \cite{goncharov99}.

In case the of metals, the use of an angular excitation geometry
was found to be crucial, since it allows a substantial reduction
in background compared to the pure back-scattering geometry. For
this, the diamond cell must be modified to allow the angular light
access. Specially designed tungsten carbide seats with angular
conical holes have been used for this purpose.

\subsection{Materials}

Samples of iron-nickel alloys Fe$_{(1-x)}$Ni${_x}$ with x$\leq$0.2
from various sources have been studied recently. Some of them
(with x=0.05, 0.10, 0.15, 0.20) were synthesized and characterized
at Cornell University \cite{huang}. Another group with x=0.05 and
x=0.1 were the same materials studied in Ref. \cite{takahashi}.
New samples with x=0.01, 0.03 and 0.07 were synthesized at the
Geophysical Laboratory. The sample with x=0.07 was enriched by
$^{57}$Fe isotope for nuclear resonance inelastic x-ray scattering
(NRIX) measurements. Rhenium samples were either commercially
available polycrystalline foils or powders in the case of our in
hydrostatic experiments. Samples of Mg$^{10}$B$_2$ were similar to
those used in Refs. \cite{bud'ko,finnemore}. They are essentially
in a powdered form and consist of aggregates of 30-50 $\mu$m
linear dimensions, which is ideal
for high-pressure experiments.

\section{Raman studies of metals}

\subsection{Iron and Fe$_{(1-x)}$Ni${_x}$ alloy}

\begin{figure}
\includegraphics[width=4in]{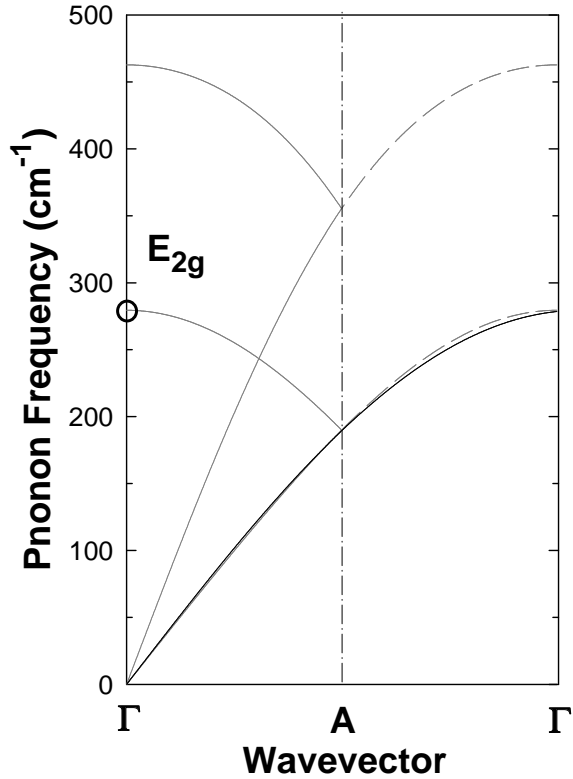}
\caption{Illustration of folding of the dispersion curves in a
hexagonal closed-packed (hcp) structure due to a doubling of the
unit cell (with respect to a cubic one). The zone boundary
transverse acoustic mode propagating along [001] direction becomes
the zone center E$_{2g}$ Raman-active mode. The gray solid and
dashed (folded) lines are first-principles calculations at about
60 GPa \protect\cite{mao2001,vocadlo_p}; the black solid line is a
"sine" curve computed with the same zone-boundary frequency (279.5
cm$^{-1}$). } \label{fig2}
\end{figure}

\begin{figure}
\includegraphics[width=4in]{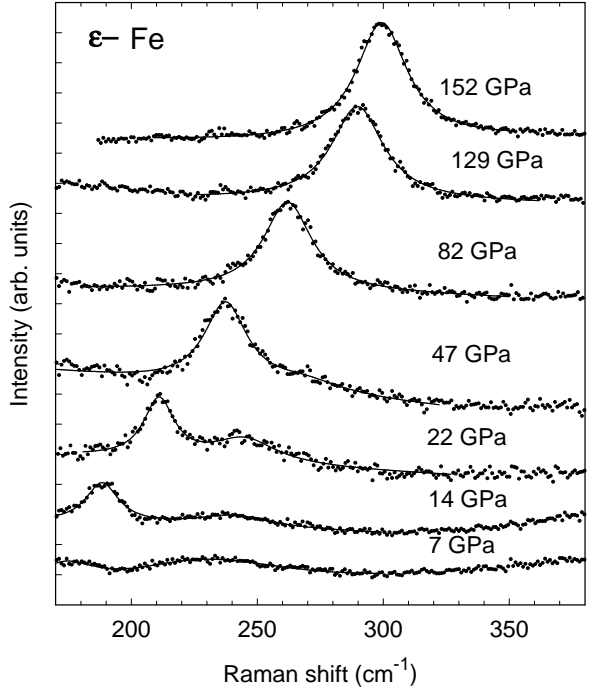}
\caption{Raman spectra of Fe at different pressures to 152 GPa
\protect\cite{merkel}. Spectra at 7 GPa and 14 GPa are measured on
pressure release. The points are experimental data; the solid
lines are phenomenological fits (Lorentzians). The spectra are
shifted vertically for clarity. } \label{fig3}
\end{figure}

\begin{figure}
\includegraphics[width=4in]{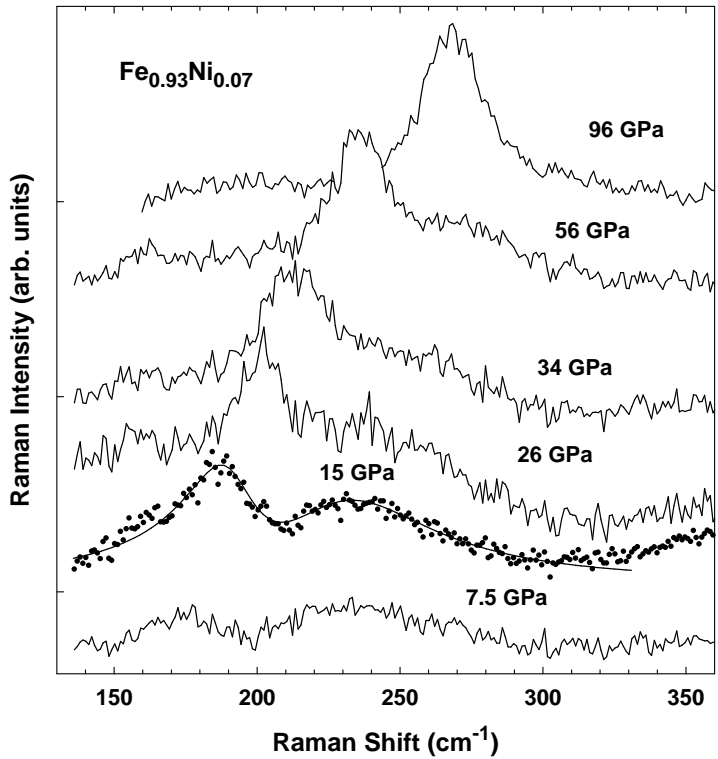}
\caption{Raman spectra of Fe$_{0.93}$Ni$_{0.07}$ at different
pressures to 60 GPa. Spectra at 7.5 GPa and 15 GPa are measured on
pressure release. The spectrum at 15 GPa is shown along with a
manual fit using a coupled oscillator model
\protect\cite{katiyar,goncharov99a,goncharovtobe}. The spectra are
shifted vertically for clarity.} \label{fig4}
\end{figure}

\begin{figure}
\includegraphics[width=4in]{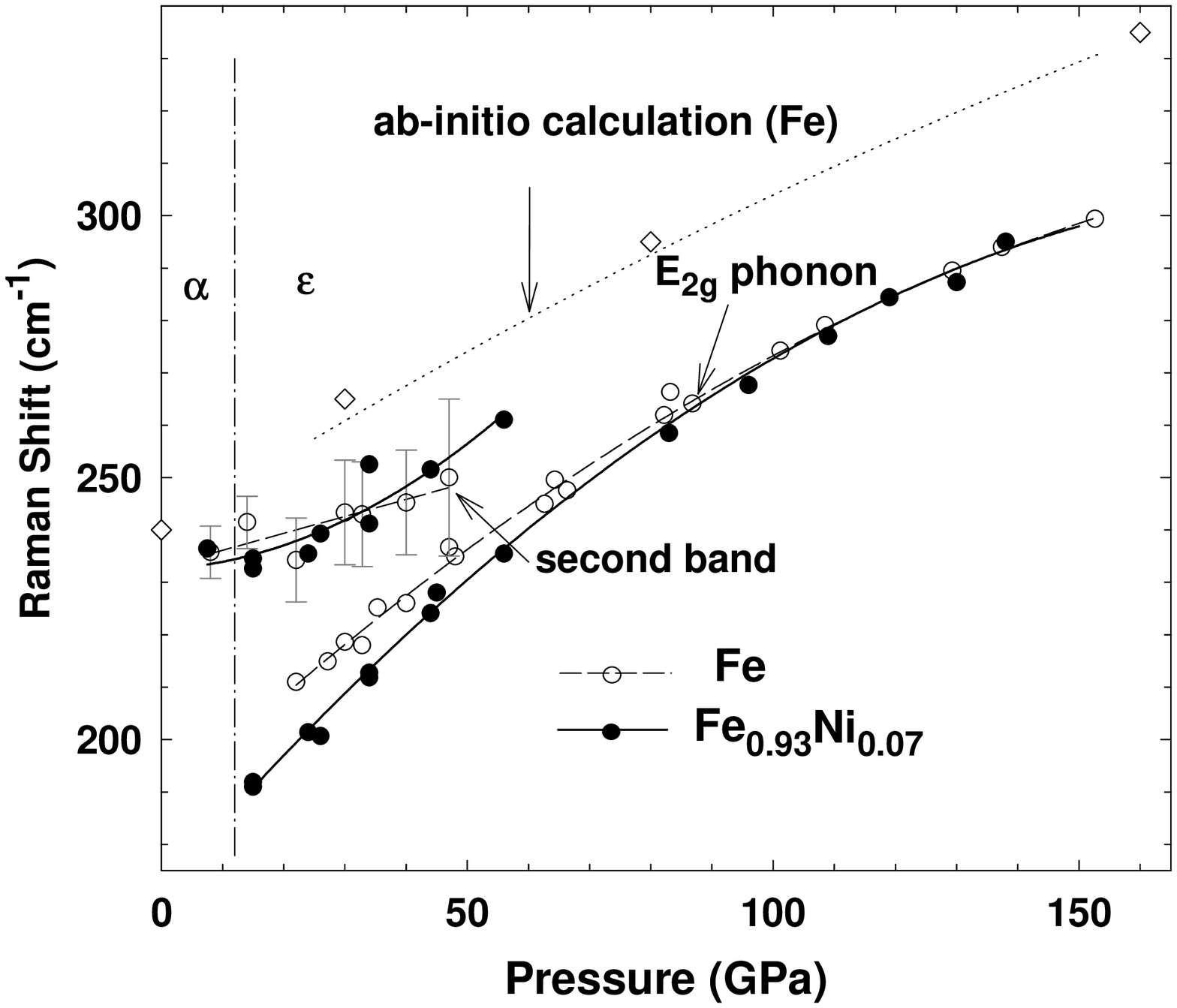}
\caption{Pressure dependence of frequencies of the Raman
excitations for pure Fe and representative Fe:Ni alloy
(Fe$_{0.93}$Ni$_{0.07}$). The curves for other x (Ni compositions)
coincide within experimental uncertainty with either that for pure
Fe (if x$<$0.03) or for the Fe$_{0.93}$Ni$_{0.07}$ compound shown
(if x$>$0.03). Circles correspond to the results of fitting
two-Voigt spectral functions (Fano plus Voigt for the 15 GPa
spectrum) to the measured data. Diamonds and dotted line are
theoretical results from Refs. \protect\cite{steinle,steinle_p}
and Refs. \protect\cite{mao2001,vocadlo_p}, respectively.}
\label{fig5}
\end{figure}

Knowledge of the physical properties of iron and its alloys is
very important for understanding the nature of the Earth's core.
Recently, there has been a flurry of experimental and theoretical
activity devoted to determination of the elastic properties of
iron at the conditions close to those of the core
\cite{stixrude,mao98,steinle,laio,lubbers,fiquet,mao2001,steinle01}.
Despite these efforts, direct and unambiguous measurements of the
second-order elasticity tensor (e.g., $C_{ij}$) under pressure are
missing, and existing experimental and theoretical data are still
controversial \cite{steinle,olijnyk01}. Raman spectroscopy can
provide independent constraints on the high-pressure elasticity
from measurements of the frequency of the transverse optical
phonon, which is related to the shear elastic constant $C_{44}$ by
the relation

\begin{equation}
C_{44}=2\pi^2M [\sqrt{3}c/(6a^2)]\nu^2,
\end{equation}

\noindent where $M$ is the atomic mass of iron, $c$ and $a$ are
unit cell parameters, and $\nu$ is the frequency of the optical
phonon \cite{olijnyk01,sharma}. This simple relation reflects the
fact that the frequency of the phonon and the sound velocity (or
corresponding $C_{ij})$ are the properties of the same phonon
branch (fig. \ref{fig2}), folded because of the doubling of the
fcc unit cell to form hcp. Figure \ref{fig2} shows the dispersion
curves in the $\Gamma$A direction (001) for the $\epsilon$-Fe
computed theoretically \cite{mao2001,vocadlo_p} in comparison with
a "sine" curve obtained with the same phonon frequency. They are
very close to each other (but a small discrepancy near the zone
boundary), which show the validity of the model used. The
theoretically computed $C_{44}$ is 232 GPa, while the model gives
a comparable value of 258 GPa. In general, this method gives
fairly good results in for metals that exhibit a "normal" behavior
under compression, while the proximity of phase transitions may
cause a substantial anisotropy of elastic properties, which makes
significantly reduced the validity of the approach (see discussion
below).

We performed the Raman measurements in $\epsilon$-Fe to 152 GPa,
which correspond to pressure at the core-mantle boundary
\cite{merkel} (see also Ref. \cite{olijnyk01} with similar
measurements to lower pressures). Raman spectra of iron at
selected pressures are shown in (fig. \ref{fig3}). The strong
band, which appears in the spectra above 15 GPa on pressure
increase and remains to approximately 10 GPa on pressure decrease,
is identified as the E$_{2g}$ mode of $\epsilon$-Fe. The
difference for compression and decompression is consistent with
the known hysteresis through the $\alpha$-$\epsilon$ transition
\cite{huang87}. A second band, which is weaker and broader than
the main one, is observed at higher frequencies at moderate
pressures. It will be discussed below in comparison with analogous
behavior found for the alloys.

We also present preliminary results of a similar study of
Fe$_{(1-x)}$Ni${_x}$ alloys and compare the results with pure Fe.
Ni can be considered a secondary element in the Earth's core, so
we addressed possible change of elastic properties of the chemical
substitution. The phase diagram of the low-nickel
Fe$_{(1-x)}$Ni${_x}$ alloys are similar to pure Fe \cite{huang},
and the transition pressures at room temperature are very close.
Moreover, the equations of state are very similar in both the
$\alpha$ or $\epsilon$ phases \cite{takahashi,mao90}. The Raman
spectra of Fe$_{0.93}$Ni$_{0.07}$ at different pressures are shown
in fig. \ref{fig4}. As for pure Fe, the dominant feature of the
spectra is the E$_{2g}$ mode, which appears when pressure exceeds
the $\alpha$--$\epsilon$ transition boundary ($>$15 GPa upstroke
in agreement with numerous x-ray data). The second broader and
weaker peak is more pronounced in the alloy compared to pure Fe
(fig. \ref{fig4}). It could be observed to almost 60 GPa (c.f., 45
GPa in pure Fe) and also after the back transformation to
$\alpha$-Fe ($<$8 GPa downstroke), where the E$_{2g}$ mode, which
is only active in $\epsilon$-Fe, becomes unobservable. The
E$_{2g}$ mode line profile appears to be asymmetric at low
pressures, which may be related to coupling to other excitations.

The frequency of the E$_{2g}$ mode is lower by some 10 cm$^{-1}$
for the alloys compared to pure Fe, and this difference decreases
with pressure, so the curves merge above 100 GPa (fig.
\ref{fig5}). Surprisingly, this Raman frequency offset is
independent of composition for x$>$0.03. For x$<$0.03 the pressure
versus frequency data are undistinguishable within the
experimental accuracy. The pressure dependence of the frequency of
the second broad Raman peak for the pure Fe and the alloys are
close to each other (fig. \ref{fig5}). In the case of the alloy,
it seems to show an upturn with pressure. This second Raman peak
was tentatively assigned in \cite{merkel} to disorder-induced
phonon scattering. According to calculations \cite{vocadlo00}, the
LA phonon branch is very flat between K and M points of the
Brillouin zone, thus giving a large density of phonon states close
in frequency to the experimentally observed second Raman peak. It
can also be due to TA zone boundary phonons at the L point of the
Brillouin zone \cite{wallace}. The large density of phonon states
in this frequency range is also confirmed by direct measurements
\cite{mao2001}. The reason for the disorder may be a proximity to
the transition boundary which can give rise to stacking faults
(mixed hcp-fcc stacking). The alloys are also naturally disordered
because of the presence of different atoms. In principle, disorder
relaxes symmetry and wave-vector conservation requirements for
Raman activity and causes coupling between excitations with
different parent symmetries. The results of a coupled oscillator
model fit \cite{katiyar,goncharov99a} are shown in fig. \ref{fig4}
\cite{goncharovtobe}. An alternative explanation would be possible
weak magnetism in $\epsilon$-Fe at lower pressures
\cite{steinle,cohen}, which would lift the degeneracy of the
E$_{2g}$ mode. Both interpretations are consistent with
disappearance of the effect at higher pressures.

The lower frequency of the E$_{2g}$ mode in the alloys compared to
pure Fe is intriguing. One would expect very small reduction of
the frequency (0.5\% or 1 cm$^{-1}$ at 25 GPa for the
Fe$_{0.8}$Ni$_{0.2}$ alloy with the maximum substitution studied)
if only the effect of changing the mass is taken into account.
Since no effect of changing of lattice parameter with Ni
substitution is reported, a substantial change in force constant
is not expected. One possible explanation for the observed
softening would be different coupling constant between the
E$_{2g}$ mode and the second Raman mode for the pure Fe and
alloys. A larger coupling constant (in case of alloys with more
relaxed selection rules, see above) would shift the E$_{2g}$ mode
to lower frequency.

The threshold nature of the effect observed (see above) suggests a
change in behavior at about x=0.03. This may be due to a change in
magnetic (or chemical) ordering or a change of free carrier
density (which in turn can influence the coupling of the E$_{2g}$
mode with free carriers). If $qv_F/h{\omega}\leq$1, where
${\omega}$ and q are the phonon frequency and wave vector, v$_F$
is the Fermi velocity, then the phonon experiences so-called
Landau damping, so that its frequency and damping re-normalize
\cite{Ipatova}. The magnitude of the effect can be tuned by
variation of pressure (through variation of ${\omega}$) and by
doping level, which change the Fermi velocity v$_F$. Additional
studies need to be performed to clarify the observed effect.

\subsection{Rhenium}

\begin{figure}
\includegraphics[width=4in]{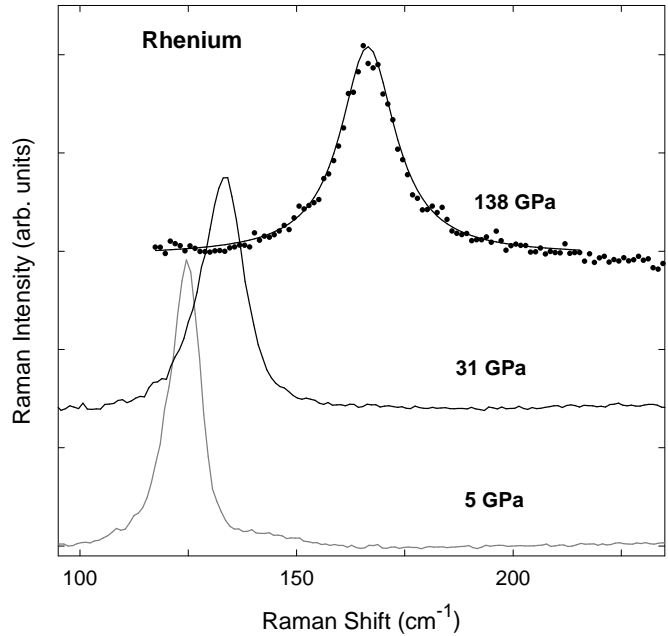}
\caption{Raman spectra of rhenium at selected pressures to 138
GPa. The spectrum at 138 GPa (dots) is shown along with Lorenzian
fit (solid line). Spectra are shifted vertically for clarity. }
\label{fig6}
\end{figure}

\begin{figure}
\includegraphics[width=4.5in]{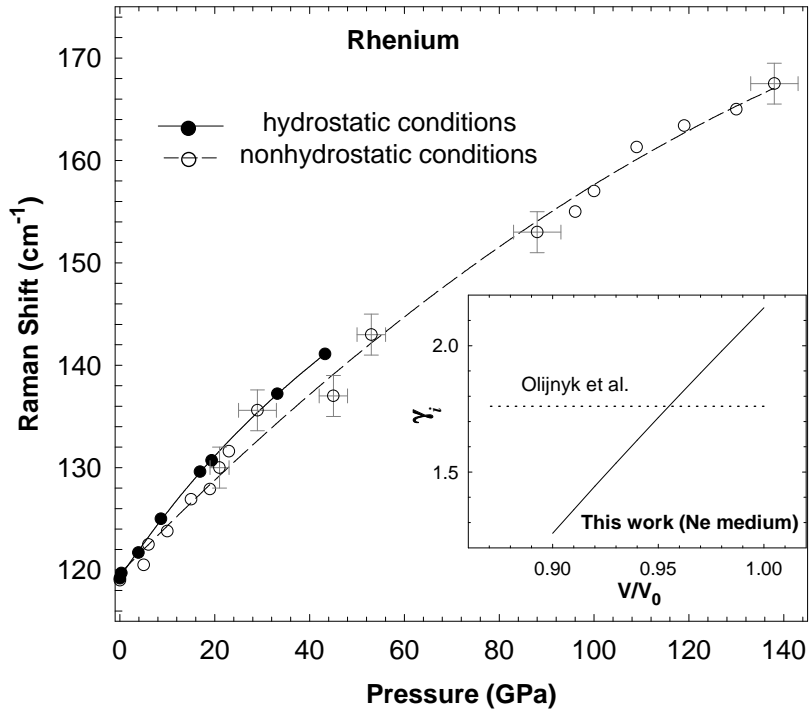}
\caption{Pressure dependence of frequency of the E$_{2g}$ Raman
phonon of Re. Filled circles - hydrostatic conditions (Ne medium),
empty symbols nonhydrostatic data. The inset shows the pressure
dependence of the mode Gr\"uneisen parameter. Solid line-our data;
dotted line -Ref. \protect\cite{olijnyk2000}.
}
\label{fig7}
\end{figure}

\begin{figure}
\includegraphics[width=4.5in]{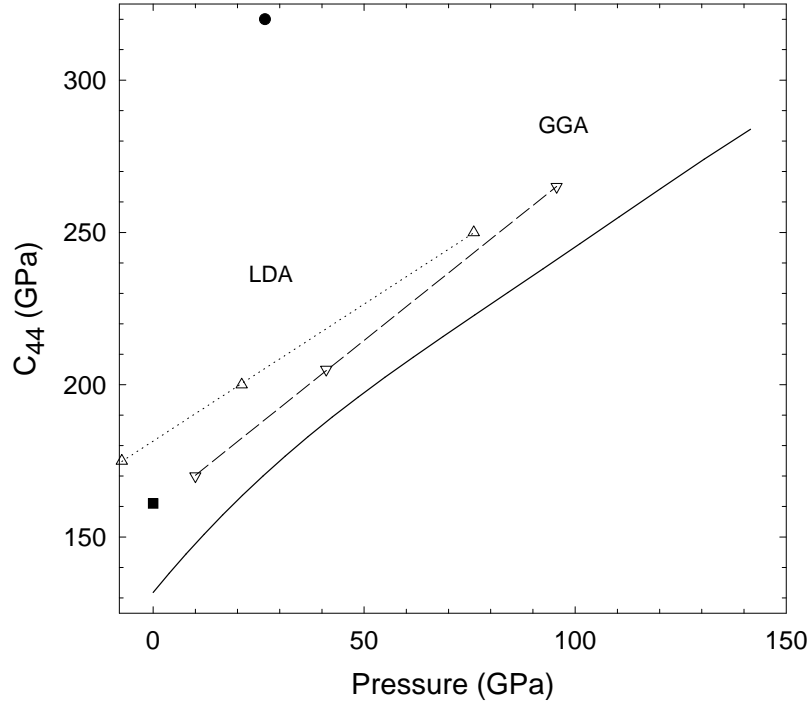}
\caption{Pressure dependence of the shear modulus $C_{44}$ of Re
deduced from the E$_{2g}$ Raman frequency. Solid line, fit to the
experimental data; squares, ultrasonic measurements
\protect\cite{manghnani}; solid circle, lattice strain
measurements \protect\cite{duffy}; upward triangles,
first-principles LDA calculations \protect\cite{steinle}; downward
triangles, first-principles GGA calculations
\protect\cite{steinle}. } \label{fig8}
\end{figure}

\begin{table}
\centering
\caption{Comparison of the measured and deduced from the Raman
measurements shear elastic modulus C$_{44}$ at ambient conditions.
}
\bigskip
    \begin{tabular*}{6.5in}{cccc}
     \hline
Element&Raman Shift(cm$^{-1}$)& $C_{44}$(GPa) calculated &
$C_{44}$
(GPa) experimental\\

\hline

  Zr  & 87  \protect\cite{olijnyk97} & 28.9  &          33.4 \protect\cite{c44}\\
  Mg  & 122  \protect\cite{olijnyk99}&  15.9   &           18.6 \protect\cite{c44}\\
  Zn  & 71.5 \protect\cite{olijnyk2000}& 19.8   &           39.6 \protect\cite{c44}\\
  Re  & 119  (this work)            &     133   &           161 \protect\cite{manghnani}\\
\hline
    \end{tabular*}
\label{table1}
\end{table}

Our initial motivation for measuring the Raman spectra of Re under
pressure was possible use it as a convenient secondary pressure
calibration gauge. As an hcp metal, Re has a Raman-active E$_{2g}$
mode, which has been measured under pressure by Jephcoat and
co-workers \cite{jephcoat,olijnyk99,olijnyk2000}; no anomalies
have been reported consistent with the wide stability field of the
phase. On the other hand, there are several peculiar observations
at ambient and low pressure, which have motivated further study.
Raman measurements at low temperatures and ambient pressure show
extra peaks, which have been ascribed to a possible structural
instability \cite{ponosov}. The pressure dependence of
superconducting transition temperature show anomalous behavior up
to 1.6 GPa \cite{chu}, which has been tentatively ascribed to a
pressure-induced change of the topology of the Fermi surface
(electronic topological transition; ETT) \cite{Lifshitz}.

We measured the Raman spectrum of Re to 138 GPa (fig. \ref{fig6}).
We found a positive monotonic shift in frequency of the E$_{2g}$
mode with pressure in agreement with the results of Ref.
\cite{olijnyk99} (which stopped at 65 GPa). A broadening of the
band with pressure (fig. \ref{fig6}) arises from nonhydrostatic
stresses because no pressure transmission medium was used (as in
case of experiments reported in Ref. \cite{olijnyk99}). The
frequency shift (fig. \ref{fig7}) is weakly sublinear with
pressure (c.f., Ref. \cite{olijnyk99}).

We also performed measurements under quasihydrostatic conditions
to 44 GPa (using Ne as a medium \cite{hemley89}) and found that
there is a systematic deviation in the frequency to the higher
energy side compared to nonhydrostatic data \cite{hanfland,xu}. As
the result, the pressure dependence of the E$_{2g}$ mode frequency
under quasihydrostatic conditions is essentially sublinear at low
pressures (fig. \ref{fig7}). The volume dependence of the
mode-Gr\"uneisen parameter (inset to fig. \ref{fig7}) obtained
from these data and x-ray diffraction measurements of the equation
of state \cite{vohra} shows a strong volume dependence in contrast
with the measurements of Olijnyk et al. \cite{olijnyk2000}. Also,
the initial mode-Gr\"uneisen parameter $\gamma_{i0}$=2.16 is
substantially larger than in Ref. \cite{olijnyk2000}
($\gamma_{i0}$=1.76). We relate the large $\gamma_{i0}$ and its
volume derivative to possible proximity of Re to ETT at $\leq$1
GPa \cite{chu} (see discussion below).

Using formalism developed in Refs. \cite{jephcoat,merkel,sharma},
we calculated the pressure dependence of shear modulus C$_{44}$
for Re from our Raman measurements of the E$_{2g}$ mode frequency
(fig. \ref{fig8}). The result at ambient pressure is 17\% lower
than the direct measurements, which appears to be a systematic
offset of the model based on comparisons among different metals
(Table \ref{table1}). This systematic deviation can be easily
explained by departure from the ideal "sine" shape of the
dispersion curve inferred in Refs. \cite{jephcoat,merkel,sharma}
because of long-range interactions. This deviation can be
substantially larger (as in the case of Zn) if there is a
substantial departure from "normal" behavior (e.g., ETT).
Nevertheless, if the validity of approximation
\cite{jephcoat,merkel,sharma} is established by an independent
method at low pressures (by direct measurements), it can be
successfully used at high pressures, which is very significant if
other measurements and/or theoretical calculations give
contradictory results.

The pressure dependence of $C_{44}$ for Re (fig. \ref{fig8})
determined in this work shows a systematic offset compared to
first-principles theoretical calculations \cite{steinle}. This
deviation is smaller at higher pressures, and it can at least
partly be accounted for by the systematic underestimation of
$C_{44}$ from the Raman frequency (see discussion below). This
result differs from the lattice strain measurements (see
discussion in Ref. \cite{duffy}).

\subsection{MgB$_2$ and ETT}

\begin{figure}
\includegraphics[width=4.0in]{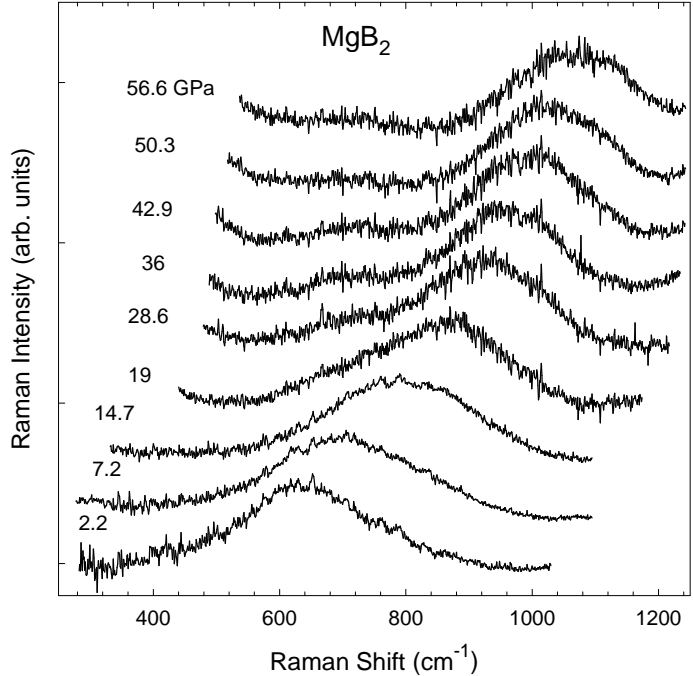}
\caption{Raman spectra of the E$_{2g}$ mode in MgB$_2$ at
different pressures. Spectra are shifted vertically for clarity.}
\label{fig9}
\end{figure}

\begin{figure}
\includegraphics[width=4.0in]{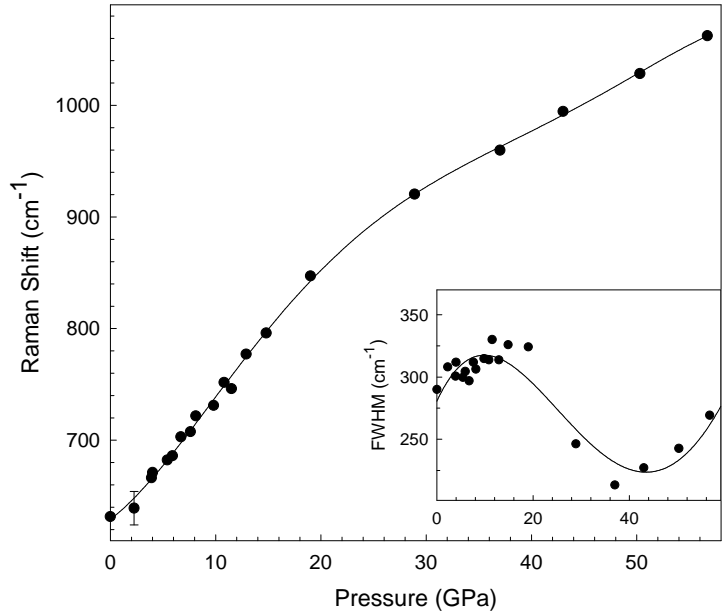}
\caption{Pressure dependence of the E$_{2g}$ Raman frequency
 \protect\cite{struzhkin01}. The inset shows the pressure
 dependence of the linewidth.
} \label{fig10}
\end{figure}

The recently discovered superconductor MgB$_2$ \cite{akimitsu}
appears to be a clear case of a conventional superconductor with
strong electron-phonon coupling
\cite{bud'ko,hinks,kortus,an,kong,yildirim,bohnen,loa}.
Theoretical calculations demonstrate that only one phonon branch
corresponding to vibrations of the B atoms perpendicular to the
$c$ axis is responsible for most of the coupling
\cite{kong,yildirim}. This branch becomes Raman-active at zone
center (E$_{2g}$ mode), which allows study of the electron-phonon
coupling by vibrational spectroscopy. Application of pressure
tunes the parameters of the system, so it can be used as an
important test for theoretical models.

A combined Raman, x-ray diffraction and $T_c$ studies under
hydrostatic pressure \cite{goncharov01,goncharov01a,struzhkin01}
to 57 GPa confirm the first-order phonon nature of the broad Raman
band and provided the volume dependence of the electron-phonon
coupling parameter. Figure \ref{fig9} shows monotonic shift of the
E$_{2g}$ mode to higher frequencies with pressure. This band is
very broad because of the electron-phonon coupling \cite{bohnen},
becoming even broader at 15-25 GPa, which may be due to the
appearance of an additional band close to the original one. At
higher pressure the band becomes narrower, and no additional Raman
band structure is observed.

The pressure dependences of the Raman frequency and linewidth of
MgB$_2$ are presented in fig. \ref{fig10}. The pressure shift of
the frequency has a kink at about 20 GPa, so the pressure
derivative of the frequency discontinuously diminishes at this
point, which corresponds to the step-like change in the
mode-Gr\"uneisen parameter (Table \ref{table2}) because no anomaly
is found in the $P-V$ curve up to 50 GPa \cite{bordet}. The
mode-Gr\"uneisen parameter from ambient to 20 GPa is anomalously
large, which we also ascribe to the electron-phonon coupling
\cite{goncharov01} and the peculiar topology of the Fermi surface
\cite{yildirim}. It has the crossing of the lower B $\sigma$ bands
with the Fermi level modulated by the zero-point motion associated
with the E$_{2g}$ mode. Pressure pushes these bands up and the
amplitude of the zero-point motion decreases with pressure,
reducing this dynamical crossing related to the E$_{2g}$ mode. It
is interesting that the $T_{c}$ versus pressure curve also changes
slope close to 20 GPa \cite{struzhkin01}.

The mode-Gr\"uneisen parameters (Table \ref{table2}) appear to be
sensitive not only to the phase transformations, but also to
changes in electronic structure (like ETT) through the
electron-phonon coupling. This is due to the fact that
$\gamma_{i}$ as well as T$_c$ are related to high-order derivative
of thermodynamic potential, and thus are more sensitive to
delicate changes in comparison to the phonon frequency or lattice
parameters. Changes in $\gamma_{i}$ ($q_{i}$) are in turn even
more sensitive and may indicate the proximity to an ETT or an
approaching instability of the lattice.

Table \ref{table2} contains information about the mode-Gr\"uneisen
parameters and their derivatives for several materials. Diamond is
known to be stable to at least 350 GPa and can be considered as an
example of "normal" behavior under pressure. Accordingly,
$\gamma_{i}$ is almost independent of pressure
$\gamma_{i}\approx$1, which is typical for covalent compounds
\cite{sherman}. Metals are expected to behave similarly; this is
illustrated by the example of $\epsilon$-Fe, which is stable to at
least 300 GPa and exhibits no substantial change of structural
parameters (e.g., $c/a$ ratio) observed \cite{mao90}. Zinc shows
anisotropic compression \cite{takemura}, and various calculations
suggest the occurrence of an ETT \cite{fast,olijnyk2000}, which
has not been confirmed experimentally
\cite{olijnyk2000,takemura99} in agreement with recent
calculations \cite{ron01}. We speculate that the enormously large
$\gamma_{i0}$ and $q_{i0}$ for Zn (and Re; fig. \ref{fig7})
indicate the proximity to an ETT, which could be at negative
pressures. With its very large electron-phonon coupling, MgB$_2$
appears to be a very clear case of an ETT driven by the phonon
\cite{struzhkin01}.

\begin{table}
\centering \caption{Comparison of initial mode-Gr\"uneisen
parameters $\gamma_{i0}$ and their volume derivatives q$_{i0}$ for
different metals}
\bigskip
    \begin{tabular*}{6.5in}{ccccc}
     \hline
Compound & $\gamma_{i0}$  & q$_{i0}$ &  Comment &  Reference \\
     \hline
diamond & 1.00(0.03) & -1(1) & normal behavior & \protect\cite{goncharov86}\\
iron ($\epsilon$ phase)& 1.68(20)& 0.7(5) & normal behavior &\protect\cite{merkel}\\
Zn  &     3    &       4.7   & close to EET  & \protect \cite{olijnyk2000}\\
Re  & 2.16(5)  &      9(3) &  close to ETT?  &  this work\\
MgB$_2$ & 2.9(3) & 0(1) & large anharmonicity &  \protect \cite{goncharov01}\\
MgB$_2$ &  1.5 (P$\geq$ 29 GPa)&   &  change of $\gamma_{i}$ is due to ETT &  \protect \cite{struzhkin01}\\
     \hline
    \end{tabular*}
\label{table2}
\end{table}

\section{Conclusions}

The development of optical spectroscopy instrumentation make it
now possible to obtain the Raman spectra of metals to megabar
($>$100 GPa) if not multimegabar pressures. The information
obtained includes the behavior of first-order phonon modes and
their coupling to the electronic and (possibly) magnetic
excitations. It can be used to estimate the elastic moduli, and to
obtain diagnostics of phase transformations and electronic
topological transitions.

\acknowledgments

The authors are grateful to G. Lapertot, S. L. Bud'ko, and P. C. Canfield
for samples of MgB$_2$. This work was supported by NSF, DOE, and the W. M. Keck
Foundation.

\end{document}